\documentclass[fleqn,usenatbib]{mnras}

\usepackage{newtxtext,newtxmath}
\usepackage{lineno} 
\usepackage[T1]{fontenc}

\DeclareRobustCommand{\VAN}[3]{#2}
\let\VANthebibliography\thebibliography
\def\thebibliography{\DeclareRobustCommand{\VAN}[3]{##3}\VANthebibliography}

\usepackage{graphicx}	% Including figure files
\usepackage{amsmath}	% Advanced maths commands

\title[SHMR at $10^{10}\,{\rm M}_\odot$ from strong lensing]{Measuring the Stellar-to-Halo Mass Relation at $\sim10^{10}$ Solar masses, \\using forthcoming space-based imaging of galaxy-galaxy strong lenses}

\author[Kaihao Wang et al.]{Kaihao Wang$^{1,2,3}$,
Xiaoyue Cao$^{4,3,1}$\thanks{E-mail: xycao@nao.cas.cn},
Ran Li$^{2,3}$,
James W. Nightingale$^{5,6,7}$,
Qiuhan He$^{6,8}$,  \newauthor
Aristeidis Amvrosiadis$^{6}$,
Richard Massey$^{6,7}$, 
Maximilian von Wietersheim-Kramsta$^{6,7}$,  \newauthor
Leo W.H. Fung$^{6,7}$,
Carlos S. Frenk$^{6}$,
Shaun Cole$^{6}$,
Andrew Robertson$^{9}$,
Samuel C. Lange$^{6}$,  \newauthor
Xianghao Ma$^{1,2,3}$
\\
$^{1}$ National Astronomical Observatories, Chinese Academy of Sciences, 20A Datun Road, Chaoyang District, Beijing 100012, China\\
$^{2}$ School of Physics and Astronomy, Beijing Normal University, Beijing 100875, China \\
$^{3}$ School of Astronomy and Space Science, University of Chinese Academy of Sciences, Beijing 100049, China\\
$^{4}$ Institute for Astrophysics, School of Physics, Zhengzhou University, Zhengzhou, 450001, China\\
$^{5}$ School of Mathematics, Statistics and Physics, Newcastle University, Newcastle upon Tyne, NE1 7RU, UK \\
$^{6}$ Institute for Computational Cosmology, Department of Physics, Durham University, South Road, Durham DH1 3LE, UK\\
$^{7}$ Centre for Extragalactic Astronomy, Department of Physics, Durham University, South Road, Durham DH1 3LE, UK \\
$^{8}$ Kapteyn Astronomical Institute, University of Groningen, PO Box 800, 9700 AV Groningen, the Netherlands \\
$^{9}$ Carnegie Observatories, 813 Santa Barbara Street, Pasadena, CA 91101, USA \\
}

\date{Accepted XXX. Received YYY; in original form ZZZ}

\pubyear{\the\year{2025}}

\begin{document}
% \linenumbers % Enable line numbers
\label{firstpage}
\pagerange{\pageref{firstpage}--\pageref{lastpage}}
\maketitle

\begin{abstract}
The stellar-to-halo mass relation (SHMR) is central to understanding the co-evolution of galaxies and their host dark matter haloes, yet it remains weakly constrained for dwarf galaxies owing to their faintness, especially beyond the Local Group. Strong gravitational lensing offers a unique probe of the SHMR at sub-galactic scales and cosmological distances, as the masses of subhalos within the main lens can be inferred from the perturbations they imprint on lensed images. Anticipating the discovery of $\sim10^5$ galaxy--galaxy strong lenses by forthcoming facilities such as \textit{Euclid}, we perform an end-to-end simulation to forecast \textit{Euclid}'s constraints on the SHMR at the halo mass scale of $\sim10^{10}\,\mathrm{M}_\odot$. We generate mock \textit{Euclid} VIS images of lens systems hosting a fiducial $3\times10^{10}\,\mathrm{M}_\odot$ subhalo and vary its properties to assess the robustness of mass inference. We find that \textit{Euclid}'s angular resolution cannot break the intrinsic mass--concentration degeneracy of subhaloes, nor deblend the light of satellite galaxies (when present) associated with them, leading to biased inferred halo masses. These limitations can be overcome with high-resolution follow-up imaging from facilities such as the \textit{Hubble Space Telescope}, enabling accurate halo-mass measurements.We forecast that a statistical sample of $\sim100$ such systems, combining lensing-derived halo masses with stellar masses from photometric SED fitting, can constrain the SHMR at dwarf-galaxy scales with a precision of $\sim0.05$~dex in halo mass and $\sim0.03$~dex in stellar mass, enabling powerful tests of galaxy formation theories.

\end{abstract}


%%%%%%%%%%%%%%%%%%%%%%%%%%%%%%%%%%%%%%%%%%%%%%%%%%

%%%%%%%%%%%%%%%%% BODY OF PAPER %%%%%%%%%%%%%%%%%%

\section{Introduction}
Within the $\Lambda$ Cold Dark Matter ($\Lambda$CDM) cosmological paradigm, cosmic structures emerge hierarchically: small dark matter haloes collapse first and subsequently grow through mergers and accretion \citep{Frenk_2012}. Baryonic gas from the cosmic web is funnelled into the haloes’ deep potential wells, where it cools, condenses and forms stars, thereby giving rise to galaxies \citep[]{White1978, Dusan2005howgalaxiesgetgas}. This intricate galaxy–halo connection is encapsulated by the stellar mass–halo mass relation (SHMR), which links a galaxy’s stellar mass ($M_{*}$) with the mass of its host halo ($M_{\rm h}$) \citep{annurev:galaxy&halo}. The SHMR provides a crucial diagnostic of the overall star formation efficiency, governed by the interplay between gas cooling and energetic feedback from supernovae and active galactic nuclei \citep[e.g.][]{1986Supernova,1998SilkQuasar,2003BensonWhatShapesLuminosity}. An accurate characterisation of the SHMR is vital not only for constraining galaxy formation theories but also for linking observable galaxy tracers to the underlying dark matter distribution—a prerequisite for testing cosmological models \citep[e.g.][]{annurev:galaxy&halo, 2003Yang, 2006Conroy}. Nevertheless, the SHMR remains poorly constrained on sub-galactic scales ($M_{*} \le 10^{9}\,{\rm M}_\odot$; $M_{\rm h} \le 10^{11}\,{\rm M}_\odot$), where galaxies are intrinsically faint and challenging to observe. Extending robust SHMR measurements into this low-mass regime is therefore imperative, as it may provide crucial insights into famous small-scale challenges--such as the missing satellites and core-cusp problems \citep{2017ARA&Asmall-scale-challenges}.

The stellar-to-halo mass relation (SHMR) of galaxies can currently be constrained through two broad classes of observational approaches: statistical and individual. Statistical techniques, most notably abundance matching (AM), infer the population-averaged relation between stellar and halo mass (together with its scatter) by connecting the galaxy stellar mass function (or luminosity function) to the theoretical halo mass function under the assumption that the most massive galaxies reside in the most massive halos (e.g., \citealt{Behroozi2010}, \citealt{Moster2012}, \citealt{Girelli2020}). In contrast, individual-based approaches first determine the stellar mass of a given system using stellar population synthesis (SPS) models \citep{2013ARA&ASPS}, and subsequently employ dynamical tracers \citep[e.g.][]{2009More, 2016SLUGGS, Reina_Campos_2022} or gravitational lensing \citep[e.g.][]{2010Auger} to estimate the mass of its host halo. Both approaches, however, encounter significant challenges in probing the SHMR at sub-galactic scales. For AM-based methods, incompleteness in the stellar mass or luminosity functions at the low-mass end hampers robust construction of the SHMR \citep[e.g.][]{Behroozi2010, Weigel_2016SMF, 2023COSMOSsmf}. For individual techniques, dynamical methods are restricted to nearby dwarf galaxies where precise kinematic measurements are feasible, precluding their application at cosmological distances. While strong gravitational lensing offers the unique capability to probe sub-galactic halos at higher redshifts ($z > 0.2$) by detecting their gravitational perturbations on extended lensed arcs, only two confirmed galaxy–galaxy strong lens systems have revealed subhalos with masses of order $10^{9}\,{\rm M}_\odot$ that can be used to constrain the SHMR (\citealt{Vegetti2010}, \citealt{Vegetti2012}), leaving the sample size too small to mitigate selection biases \citep{Amorisco2022} or provide statistically powerful constraints. As such, while strong lensing remains a theoretically promising avenue for exploring the SHMR at cosmological scales, its practical implementation awaits the discovery and characterisation of a substantially larger sample of suitable lens systems.

Fortunately, with the advent of a new generation of space-based survey telescopes such as Euclid \citep{EuclidOverview}, CSST \citep{csstcollaboration2025introductionchinesespacestation}, and Roman \citep{Roman}, hundreds of thousands of galaxy–galaxy strong lenses are expected to be discovered \citep[e.g.][]{cao2023csststronglensingpreparation}, potentially yielding thousands of detections of subhalos with masses of order $10^{10}\,{\rm M}_\odot$ \citep[e.g.][]{Euclid_sensitivity}. This increase in sample size alleviates the limitation imposed by the current paucity of lenses, making strong lensing a practical tool for constraining the SHMR on sub-galactic scales. However, although previous studies have demonstrated that the Hubble Space Telescope (HST) can model the masses of $\sim\!10^{9}\,{\rm M}_\odot$ subhalos without significant bias \citep{Vegetti2009, He2023}, it remains unclear whether survey telescopes such as Euclid and CSST—whose lensing observations may have lower angular resolution and shallower depth—can achieve comparable performance. Moreover, \citet{Amorisco2022} showed that highly concentrated subhalos can produce lensing perturbations similar to those induced by more massive but less concentrated subhalos. It is therefore important to assess how this mass-concentration degeneracy impacts the accurate measurement of subhalo masses.

In this study, we use a suite of mock strong-lensing observations to quantify the accuracy and potential biases in measuring the masses of subhalos with $M_{\rm h} \sim 10^{10}\,\mathrm{M}_{\odot}$, with the aim of evaluating the constraining power of strong lensing on the SHMR at sub-galactic scales. This $M_{\rm h} \sim 10^{10}\,\mathrm{M}_{\odot}$ mass scale is chosen because it lies within the typical detection capability of the new generation of space-based survey telescopes, such as Euclid and CSST \citep[see, e.g.][]{Euclid_sensitivity}. As a proof of concept, we adopt a fiducial subhalo mass of $M_{\rm h} = 3 \times 10^{10}\,\mathrm{M}_{\odot}$ and systematically investigate how the mass measurement is affected by (i) the subhalo’s intrinsic properties, such as its concentration and position, and (ii) the baryonic components of its hosted satellite galaxy, including both stellar mass and light. Mock lensing images are generated to match the specifications of the Visible Camera (VIS) on Euclid. Since the CSST delivers image quality comparable to Euclid's VIS, the main results of this study are expected to be transferable to CSST observations.

The structure of this paper is as follows: In Section~\ref{sec:method}, we describe the mock simulations and fitting procedures. Section~\ref{sec:results} presents our lens modelling results. In Section~\ref{sec:discussion}, we discuss the implications of our results and consider prospects for future applications. We summarise our findings in Section~\ref{sec:conclusion}. We adopt the cosmological parameters from Planck 2015 \citep{Planck2015} throughout the work.

\section{Method}
\label{sec:method}
For this proof-of-concept study, we perform an end-to-end analysis of simulated galaxy-galaxy strong lens systems. Our methodology is designed to assess how accurately the properties of a low-mass subhalo can be recovered from data with the quality expected from the Euclid VIS instrument. We focus on a single, idealised lensing configuration with a prominent Einstein ring, as such systems provide the highest sensitivity for subhalo detection\footnote{As demonstrated by \citep{Despali2022, Euclid_sensitivity}, lenses with ring-like morphologies generally provide higher sensitivity for subhalo detection. Consequently, it is expected that most ``golden lens'' samples identified in future surveys, such as Euclid and CSST, which enable accurate subhalo detection, will exhibit ring-like structures. Thus, our fiducial lensing image serves as a representative morphology for evaluating the accuracy of subhalo detection. Space-based imaging surveys like CSST and Euclid are anticipated to discover tens of thousands of ring-like lenses \cite[e.g.][]{cao2023csststronglensingpreparation}, making the exploration conducted in this work both realistic and most relevant for future studies.}. 
We first generate mock images of this lens system and embed a subhalo with a fiducial mass of $\sim 10^{10.5}\,\mathrm{M}_{\odot}$. We then model these images using the open-source software \texttt{PyAutoLens} \citep{Nightingale2015, Nightingale_2018, pyautolens} to measure the subhalo mass and quantify the precision with which the stellar-to-halo mass relation (SHMR) can be constrained. We vary the subhalo's position and concentration in the mock image to test the robustness of the modelling. We also introduce components representing the mass and light of a potential satellite galaxy hosted by the subhalo to investigate their impact on the mass measurement.

\subsection{Image Simulation}
\label{sec:model} % used for referring to this section from elsewhere 
\begin{figure}
    \centering
    \includegraphics[width=\columnwidth]{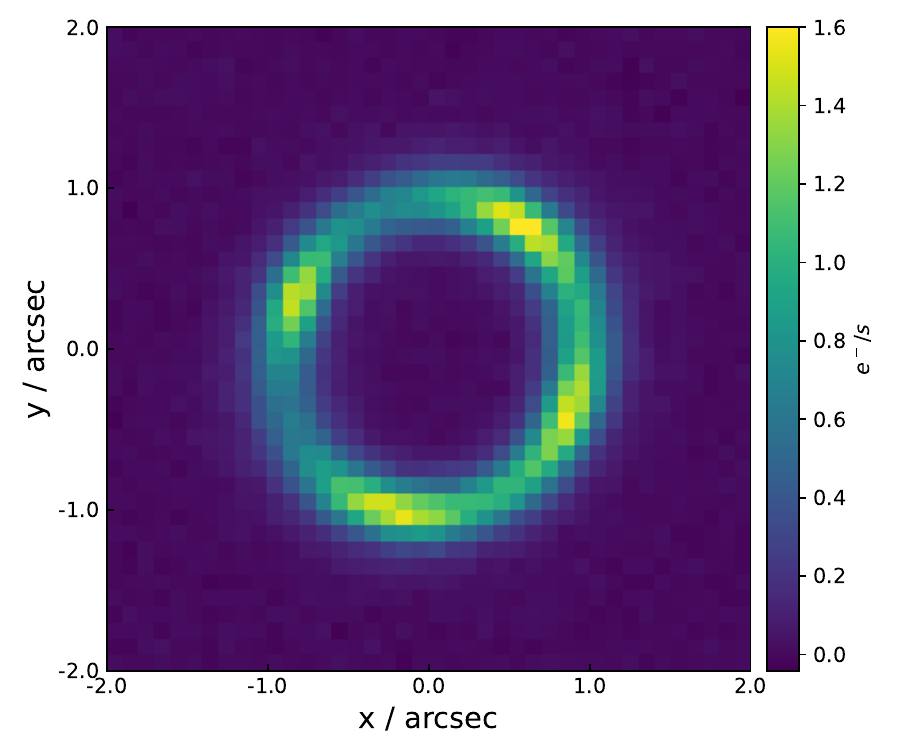}
    \caption{The simulated mock image based on the macro model. The parameter values used for this simulation are listed in Table~\ref{tab:param}. The image has a pixel size of $0.1\arcsec$ and is convolved with a Gaussian PSF with a FWHM of $0.18\arcsec$.}
    \label{fig:mock}
\end{figure}

Our simulations begin with a baseline ``macro model'' that includes only the main lens galaxy and the background source. All subsequent simulations that incorporate additional subhalos are built upon this macro model. Figure~\ref{fig:mock} shows the lensed image generated from this baseline model.

\subsubsection{Macro Model}
\label{sec:macro}
The mass distribution of the main lens galaxy is described by an elliptical power-law profile \citep{Tessore2015}. Following \citet{Nightingale2024}, the convergence is given by:
\begin{equation}
    \kappa(\xi)=\frac{3-\gamma}{1+q}\left(\frac{R_E}{\xi}\right)^{\gamma-1},
	\label{eq:power-law}
\end{equation}
where $\gamma$ is the density slope, $R_E$ is Einstein radius, and $\xi=\sqrt{x^2+y^2/q^2}$ is the elliptical radius from centre. In \texttt{PyAutoLens} the axis ratio $q$ and the position angle $\phi$ are reparameterized as two ellipticity components:
\begin{equation}
    e_1=\frac{1-q}{1+q}\sin{2\phi}, \quad
    e_2=\frac{1-q}{1+q}\cos{2\phi}.
    \label{eq:axis_ratio}
\end{equation}

Although external shear \citep{Witt_1997} is not included in the simulation of mock images, it is incorporated during the lens modelling process. This inclusion is essential because external shear is a standard component in lens modelling, accounting for line-of-sight effects, nearby perturbations, and unmodelled angular complexities in the main lens mass distribution \citep{slam_Cao_2022, Etherington_Shear}. Moreover, external shear can partially absorb the subhalo signal; thus, ignoring it in lens modelling would lead to an overestimation of subhalo detection sensitivity. The brightness distribution of the background source galaxy is described by a cored S\'ersic profile \citep{Core_sersic}, which is defined as\footnote{The source morphology can also influence the detectability of a subhalo \citep{hughes2024impactssourcemorphologydetectability}. Irregular sources with steeper spatial brightness gradients generally yield higher subhalo-detection sensitivity. As the primary aim of this work is to assess how changes in the subhalo’s own properties (e.g. position, concentration) affect the accuracy of mass measurements for \textit{detected} subhalos, we therefore adopt simplified core--S\'ersic source models throughout.}
\begin{equation}
    I(\xi) = I' \left(1 + \left(\frac{r_b}{\xi}\right)^\alpha\right)^{\frac{\gamma}{\alpha}} \exp\left[-b_n \cdot \left(\frac{\xi^\alpha + r_b^\alpha}{r_e^\alpha}\right)^{\frac{1}{\alpha n}}\right],
    \label{eq:core_sersic}
\end{equation}
where $I'$ is the scaled intensity factor, $r_b$ is the break radius separating the inner power-law and outer S\'ersic profile, $\alpha$ controls the transition from the inner to outer profile, $\gamma$ is the power-law density slope, $n$ is the S\'ersic index, $b_n$ is a constant determined by $n$, and $r_e$ is the effective radius of the S\'ersic profile. The radius $r$ is measured from the centre, and for an elliptical profile, it is defined as $r = \sqrt{qx^2 + y^2/q}$.

\subsubsection{Subhalo and Associated Galaxy}
\label{sec:subhalo}
We adopt the Navarro-Frenk-White (NFW) mass profile \citep{NFW} for subhalo:
\begin{equation}
    \rho(r) = \frac{\rho_0}{(r/R_s) (1 + r/R_s)^2}.
\end{equation}

In \texttt{PyAutoLens}, the virial mass $M_{200}$\footnote{In this work, we adopt the virial mass $M_{200}$ as a proxy for the halo mass $M_{\rm h}$. Throughout, the symbols $M_{200}$ and $M_{\rm h}$ are used interchangeably to denote the same quantity.} and the concentration $c$ are supplied as inputs to compute the characteristic density $\rho_0$ and the scale radius $R_s$. Here, $M_{200}$ denotes the mass enclosed within the radius $r_{200}$ at which the mean enclosed density equals 200 times the cosmological critical density. Throughout this work, we adopt a fiducial mass of $M_{200} = 3 \times 10^{10}\,\mathrm{M}_\odot$. The concentration is defined as $c \equiv r_{200}/R_{\mathrm{s}}$ and exhibits a well-established dependence on halo mass and redshift—often termed the mass–concentration–redshift relation—denoted $c(M,z)$ \citep[e.g.][]{mcr_Duffy,mcr_Ludlow}. \citet{Wang2020} used large-volume hierarchical simulations to show that the scatter in halo concentration at fixed mass is nearly independent of mass, with a value of approximately $0.15$ dex. Throughout, we adopt the mass-concentration relation of \citet[][hereafter L16]{mcr_Ludlow}, as it shows the best agreement with recent simulations in the very low-mass regime \citep{Wang2020}. We allow for intrinsic scatter around this relation by parameterising the concentration in terms of its normalised deviation, $s_{\mathrm{c}}$, from the median L16 relation:
\begin{equation}
s_{\mathrm{c}} \equiv \frac{\delta \log_{10} c}{\sigma_{\log_{10} c}}
= \frac{\log_{10} c - \log_{10} c(M,z)}{\sigma_{\log_{10} c}},
\end{equation}
where $\sigma_{\log_{10} c} = 0.15$. By default, we set $s_{\mathrm{c}}=0$, except when we vary this parameter to assess the impact of concentration.

A subhalo of mass $10^{10}\,\mathrm{M}_\odot$ is expected to host a dwarf galaxy. Although predictions for the stellar mass at fixed halo mass exhibit substantial scatter, a $10^{10}\,\mathrm{M}_\odot$ subhalo typically hosts a satellite with a stellar mass of $\sim 10^{8}\,\mathrm{M}_\odot$ or less \citep{Girelli2020}. To remain conservative, we adopt the stellar–halo mass relation of \citet{Girelli2020}, which yields a stellar mass of $10^{8.22}\,\mathrm{M}_\odot$—larger than most alternative predictions. Following \citet{0946_Ballard}, we assume a stellar mass-to-light ratio of $2\,\mathrm{M}_\odot/\mathrm{L}_\odot$, corresponding to an absolute magnitude of $-15.19$ (equivalently, an apparent magnitude of $24.83$ at $z=0.2$). These dwarf‐galaxy properties are used to generate mock lenses that include the light and mass contributions of the satellite galaxy discussed in Section~\ref{sec:galaxy}.

\subsubsection{Instrument information and data quality}
\label{sec:instrument}
The mock images are simulated to match the specifications of the Euclid VIS instrument. We assume a total exposure time of 1695 seconds, corresponding to a typical three-exposure observation in the Euclid survey. The images have a pixel scale of $0.1\arcsec$ and are convolved with a Gaussian point spread function (PSF) with a full width at half maximum (FWHM) of $0.18\arcsec$. The magnitude zero point of VIS is 25.23 \citep[in electrons;][]{preparationIV}. Combined with a sky background surface brightness of $22.2 \, \mathrm{mag/arcsec^2}$ \citep{Collet2015}, this corresponds to a flux of $0.1629 \, \mathrm{e^- s^{-1} pixel^{-1}}$. The readout noise of the Euclid VIS instrument is $<4.5 \, \mathrm{e^-}$, which is negligible compared to other noise sources and is therefore excluded from this analysis. The apparent magnitude of the source before lensing is set to be 26.84, which is a representative value for the galaxy-galaxy strong lensing catalogue by \citet{Collet2015}. This setup results in a maximum pixel signal-to-noise ratio (SNR) of approximately 50 in the brightest parts of the lensed arc. Table~\ref{tab:param} gives a summary of the model parameters used in this simulation.

\begin{table}
 \caption{Model components used to generate the mock lensing image. Parameter values are randomly selected from physically plausible ranges.}
 \label{tab:param}
 \begin{tabular}{lcc}
  \hline
   & values\\
  \hline
  \textbf{Input lens mass} & \\
  \hline
  \textbf{model name} & \textbf{elliptical power-law} \\
  centre (x, y) [($\arcsec$, $\arcsec$)] & (-0.0012, -0.0342) \\
  axis ratio & 0.7 \\
  position angle [$^\circ$] & 156.6 \\
  Einstein radius [$\arcsec$] & 1.0 \\
  slope & 2.16 \\
  redshift & 0.2 \\
  mass within Einstein radius [${\rm M}_\odot$] & $1.35\times10^{11}$ \\
  \hline
  \textbf{Input subhalo mass} & \\
  \hline
  \textbf{model name} & \textbf{NFW} \\
  $M_{200}$ [${\rm M}_\odot$]& $3\times10^{10}$ \\
  redshift & 0.2 \\
  \hline
  \textbf{Input satellite galaxy (if added)} & \\
  \hline
  \textbf{model name} & \textbf{spherical S\'ersic} \\
  intensity (light) [$e^-\,s^{-1}\,pixel^{-1}$] & $7.35\times10^{-3}$ \\
  intensity (convergence) [$\kappa$] & $2.33\times10^{-3}$ \\
  $r_e$ [$\arcsec$] & 0.38 \\
  S\'ersic index & 1.3 \\
  \hline
  \textbf{Input Source light} & \\
  \hline
  \textbf{model name} & \textbf{core S\'ersic} \\
  intensity at $r_b$ [$e^-\,s^{-1}\,pixel^{-1}$] & 1.919 \\
  centre (x, y) [($\arcsec$, $\arcsec$)] & (0.03, -0.05) \\
  ($e_1$, $e_2$) & (0.073, 0.061) \\
  $r_e$ [$\arcsec$] & 0.18 \\
  $r_b$ [$\arcsec$] & 0.05 \\
  S\'ersic index & 1.7 \\
  $\alpha$ & 2.0 \\
  $\gamma$ & 0.0 \\
  redshift & 1.0 \\
  \hline
 \end{tabular}
\end{table}

\subsection{Lens Modeling}

In this study, the model used to fit the mock images is nearly identical to that employed in our simulations, with two exceptions. First, although the simulated source has a parametric profile, we model it on a flexible, pixelated grid. Second, as noted in Section~\ref{sec:macro}, we include an external shear component in the model, even though it was not used when generating the mock images. We adopt this modelling choice not only because a pixelated source and external shear are standard in contemporary lensing analyses, but also because both components can partially absorb the signal of a subhalo. Omitting them could therefore overstate the inferred detectability of subhalos when extrapolating our results to real observations. This section outlines the source reconstruction method and the model fitting procedure.

\subsubsection{Adaptive Source Reconstruction}
\label{Source}

For a given set of mass model parameters, the lensed source emission is ray-traced back to the source plane and reconstructed onto an adaptive mesh. Following the formalism of \citet{Nightingale2024}, we model the source using a Voronoi mesh with natural neighbor interpolation \citep{Sibson1981} and apply adaptive regularization to smooth the reconstruction based on the source luminosity at each point\footnote{In {\tt PyAutoLens}, these schemes correspond to the {\tt VoronoiNNBrightnessImage} pixelization and the {\tt AdaptiveBrightnessSplit} regularization.}. The likelihood function for these fits follows \citet{Suyu2006} and is provided in equation 17 of \citet{Nightingale2024}. Appendix A of \citet{He2024} also illustrates the Voronoi natural neighbour interpolation scheme and adaptive regularisation in detail.

A key step in the source analysis is defining the centres of the source pixels. Previous {\tt PyAutoLens} works have employed two approaches. The first overlays a rectangular Cartesian grid over the image plane, tracing all coordinates to the source plane. The second uses a weighted K-Means clustering algorithm \citep{scikit-learn} to distribute image-plane coordinates such that they cluster around the brightest regions of the lensed source galaxy. The first method adapts to the \textit{lens mass magnification}, while the second adapts to both the \textit{lens mass magnification} and the \textit{lensed source emission}. However, the K-Means algorithm is limited as it cannot assign more than one source pixel per image pixel. This constraint limits the source reconstruction resolution and can introduce bias, which is a significant concern for the lower-resolution telescopes such as Euclid. A straightforward solution is to apply KMeans to an over-sampled image (e.g., dividing each image pixel into $11\times11$ sub-pixels), allowing multiple source pixels to cluster within one image pixel \citep[e.g.][]{Minor2024}. However, it is time-consuming to apply the K-Means algorithm directly to such a large 2D array. To address this, we have developed a more efficient clustering scheme. Our method first converts the 2D image into a 1D array using a Hilbert space-filling curve \citep{Hilbert1891}. This curve is weighted according to the flux on each pixel so it closely traces the brightest regions of the image. Then we probabilistically draw $N$ $(y, x)$ points from this Hilbert curve. The majority of the points will be drawn from high-weighted regions, thereby tracing the brightest regions of the image. These points are then traced back to the source plane via the mass model to set up the source pixel centres. In this way, we are able to map more source pixels to the brightest regions of where the source is actually reconstructed. The clustering strength can be controlled by rescaling the flux distribution of the lensed source. The new clustering scheme is named {\tt HilbertMesh} in {\tt PyAutoLens}, and a more detailed description is provided by He et al. in prep. (2026).

\subsubsection{Fitting procedure}
\label{sec:fitting}
We use the \texttt{SLaM} pipelines implemented in \texttt{PyAutoLens} to model the lensing images \citep[see also][]{slam_Amy22, slam_Cao_2022, He2023, Nightingale2024}. These pipelines employ a non-linear search chaining method, which breaks a high-dimensional, complex parameter space into a series of simpler, more manageable searches. Initial searches use simplified models and explore less complex parameter spaces, establishing priors for subsequent, more detailed searches. This ensures that the final, comprehensive search begins near the global maximum. This sequential approach improves the efficiency of the sampling process and reduces the risk of missing the global best solutions. In this work, we do not simulate lens light, so no model for the lens light is included in the fit. During the fitting procedure, non-linear searches are performed using the neural network-based sampler \texttt{Nautilus} \citep{nautilus}. This sampler uses nested sampling, returning Bayesian evidence which is essential for subhalo detection. The full subhalo modelling pipelines are as follows:

\textbf{(a) \textit{Source Parametric Pipeline}}: This initial pipeline fits the lensing image with a simple parametric source model to provide a rapid estimate of the lens mass parameters. The lens mass distribution is parameterised by a Singular Isothermal Ellipsoid (SIE) profile with external shear, while the source light is modelled using a cored-S\'ersic profile.

\textbf{(b) \textit{Source Pixelised Pipeline}}: To reconstruct the complex structure of the background source, this pipeline employs a pixelised source model based on a Voronoi mesh with adaptive regularisation. The procedure comprises two consecutive fits. In the first fit, the centres of the source Voronoi mesh are determined by overlaying a Cartesian grid on the image plane and mapping its coordinates to the source plane via the lens mass model. The lens mass parameters, initialised from the \textit{Source Parametric Pipeline} using Gaussian priors, are optimised jointly with the pixelised source regularisation. This step produces a refined estimate of the lens mass and a robust characterisation of the lensed source morphology. In the second fit, the lens mass parameters are fixed to the values derived in the first fit, and the model image of the lensed source from the first fit is used to construct an adaptive pixelisation that allocates more pixels to brighter source regions. This enhances the source reconstruction resolution and improves the quality of the lens modelling fit.

\textbf{(c) \textit{Mass Pipeline}}: After establishing the optimal pixelisation scheme for the source light, we employ a more advanced mass model, parameterised by an elliptical power-law profile with external shear.

\textbf{(d) \textit{Subhalo Pipeline}}: This final pipeline searches for a dark-matter subhalo by comparing the Bayesian evidence of lens models with and without a subhalo. First, to estimate precisely the evidence for the no-subhalo model, we refit the lensing image using the settings of the final fit from the mass pipeline, with the mass-parameter priors also anchored to that fit. For the subhalo model, the parameter space is complex and the perturbations induced by a subhalo are subtle, making direct sampling inefficient and unreliable. Instead, we adopt a grid-search strategy, first partitioning the image into a $6\times 6$ grid of cells. Within each cell, we fit a lens model that includes a candidate subhalo whose position is drawn from a uniform prior over that cell. This procedure yields $6\times 6$ individual fits and an associated $6\times 6$ map of the evidence increase relative to the no-subhalo model; we then identify the cell with the largest increase in evidence as the most probable location of the candidate subhalo. Using this location from the grid-search, we perform a final, refined fit in which the subhalo position is sampled from a Gaussian prior centred on the most-probable location with a standard deviation of $0.5\arcsec$.
The $M_{200}$ is assigned a log-uniform prior from $10^6{\rm M}_\odot$ to $10^{12}{\rm M}_\odot$. This fit provides the final Bayesian evidence for the model with a subhalo. If a subhalo is present, a significant increase in Bayesian evidence is expected compared to the no-subhalo model. According to Jeffreys' scale \citep{jeffreys1998theory}, fits with $\Delta\ln{E} \geq 5$ are classified as detections, where $\Delta\ln{E}$ represents the increase in the logarithm of Bayesian evidence from the no-subhalo model to the subhalo-included model, and $\Delta\ln{E} = 5$ corresponds to a significance of $3.6\sigma$.

\begin{figure}
    \centering
    \includegraphics[width=\columnwidth]{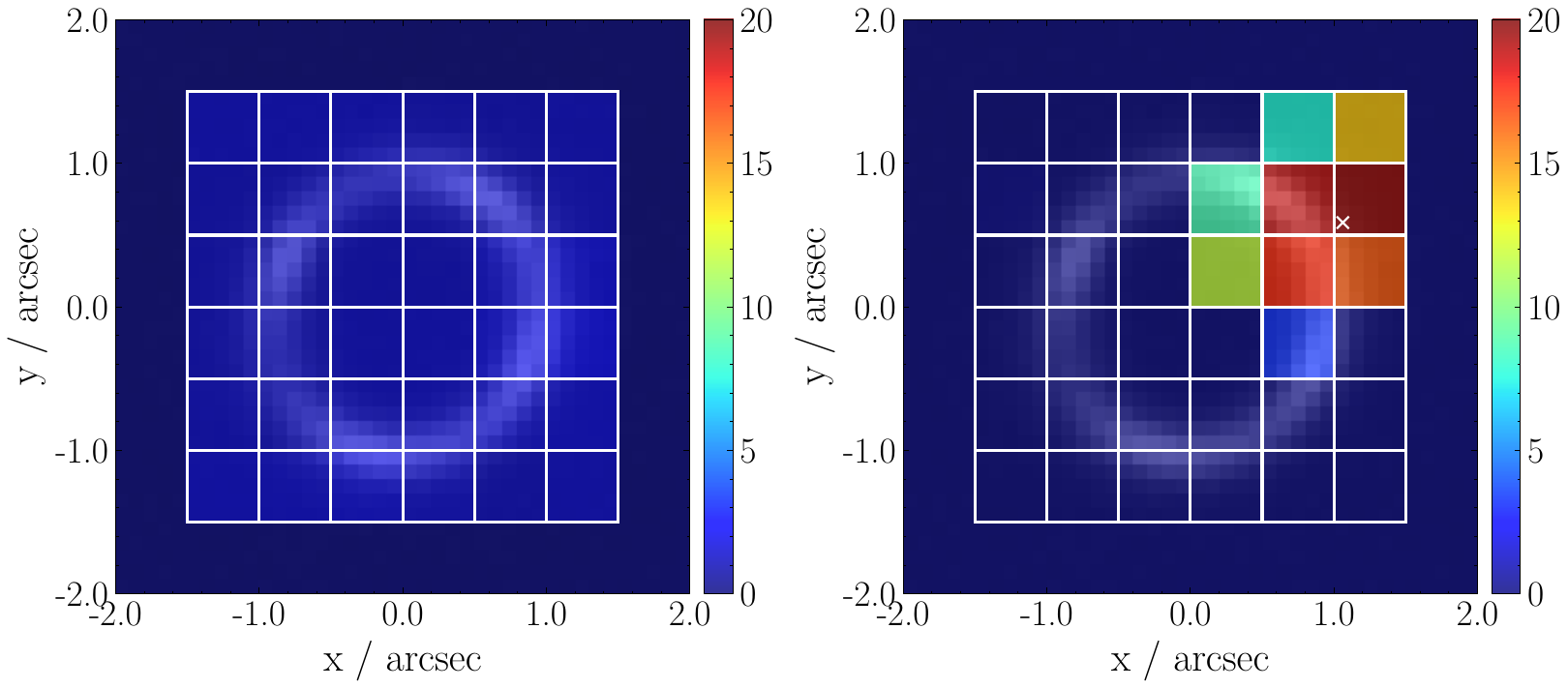}
    \caption{Evidence--increase maps for two mock lens systems illustrating the subhalo grid-search strategy. The colour bar shows the increase in logarithmic evidence, $\Delta\ln E$, obtained when a subhalo is included in the lens model with its position restricted to each $0.5\arcsec \times 0.5\arcsec$ grid cell. \textbf{Left:} A system with no subhalo. No grid cell shows a significant evidence increase ($\Delta\ln E < 5$), demonstrating that the method does not produce false positives. \textbf{Right:} A system with an injected subhalo. The grid cell with the maximum evidence increase of $\Delta\ln E = 19.55$ correctly identifies the most probable location of the subhalo.
    }
    \label{fig:example_detection}
\end{figure}

In Figure~\ref{fig:example_detection}, we show evidence increase maps from the grid-search step for two mock lens images to illustrate the subhalo detection procedure. The system on the left contains no input subhalo, whereas the system on the right includes an input subhalo whose position is marked by a white cross. For each $0.5\arcsec \times 0.5\arcsec$ cell, the colour bar encodes the evidence increase, $\Delta\ln E$, from the best grid-search result within that cell. In the left panel, no grid cell has $\Delta\ln E > 5$, indicating that no subhalo is detected. This is expected, as no subhalo was injected into the mock image, demonstrating that our modelling does not produce false positives. In the right panel, the peak evidence increase is $\Delta\ln E \simeq 19.6$, which is well above the detection threshold of 5; we therefore report a probable subhalo detection in this image. This detection would typically be validated by refining the model with a more flexible positional prior (e.g., a broad Gaussian) instead of confining the subhalo to a single grid cell. While this refinement can slightly lower the final $\Delta\ln E$ value, the change is generally negligible and does not affect the outcome of the detection.

\section{Results}
\label{sec:results}
We present the results of our lens modelling analysis. We first assess the accuracy and precision with which subhalo mass can be measured from Euclid-quality data in Section~\ref{sec:position}. We then investigate potential systematic biases introduced by the satellite galaxy hosted within the subhalo in Section~\ref{sec:galaxy}. Finally, we forecast the constraints that a sample of such detections could place on the SHMR at a mass scale of $\sim 10^{10} {\rm M}_\odot$ in Section~\ref{sec:constrain}.

\subsection{Measuring subhalo properties}
\label{sec:position}
We simulate 161 mock images, each containing a single subhalo uniformly distributed within an annulus with inner and outer radii of 0.55\arcsec and 1.5\arcsec, respectively. The input subhalo mass is fixed at $3\times10^{10}\,{\rm M}_\odot$ with concentration set by the L16 relation. To assess the accuracy with which the subhalo mass is recovered, we define an error coefficient $\mathcal{B}$ via
\begin{equation}
\label{eq:bias}
\frac{1}{2}\bigl[1+\operatorname{erf}\!\bigl(-\tfrac{\mathcal{B}}{\sqrt{2}}\bigr)\bigr] = \Phi_P\!\bigl(m_{200,\mathrm{true}}\bigr),
\end{equation}
where the left-hand side is the cumulative distribution function (CDF) of a standard normal distribution evaluated at $\mathcal{B}$, and the right-hand side is the posterior CDF of $m_{200}$ produced by the lens model, evaluated at the true input value (i.e., $3\times10^{10}\,{\rm M}_\odot$). Thus, $\mathcal{B}$ quantifies the posterior credibility at which the true mass is recovered; smaller $\lvert\mathcal{B}\rvert$ indicates closer agreement between the inferred and true values. In the absence of systematic modelling errors, $\mathcal{B}$ should follow a standard normal distribution, implying that $99.7\%$ of realisations satisfy $\lvert\mathcal{B}\rvert \le 3$.

\begin{figure}
    \centering
    \includegraphics[width=\columnwidth]{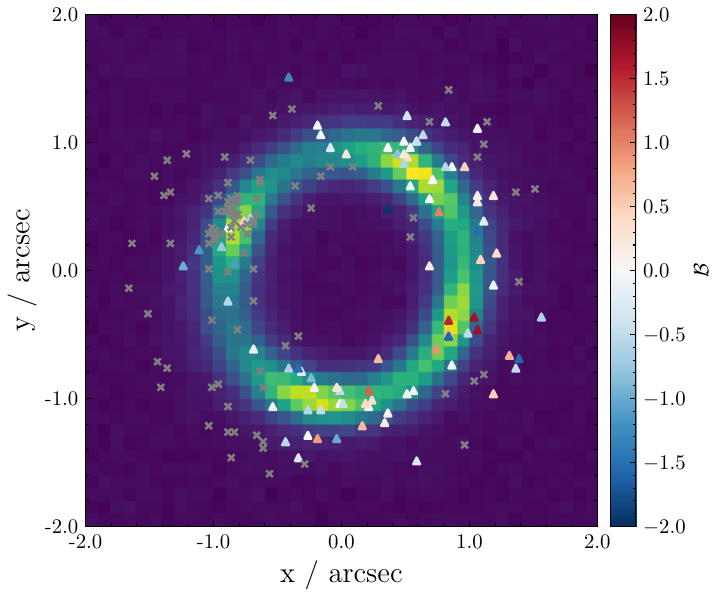}
    \caption{Accuracy of subhalo mass measurements for a suite of mock lenses at various input positions. The background image displays the lensing morphology generated from the macro model. Triangles indicate the locations of successfully detected subhalos, with their colour representing the mass error coefficient $\mathcal{B}$. This coefficient reflects the difference between the inferred posterior and the true value, defined following equation~\ref{eq:bias}. Grey crosses mark locations where the input subhalo was not detected by our pipeline.}
    \label{fig:bias_position}
\end{figure}

\begin{figure}
    \centering
    \includegraphics[width=\columnwidth]{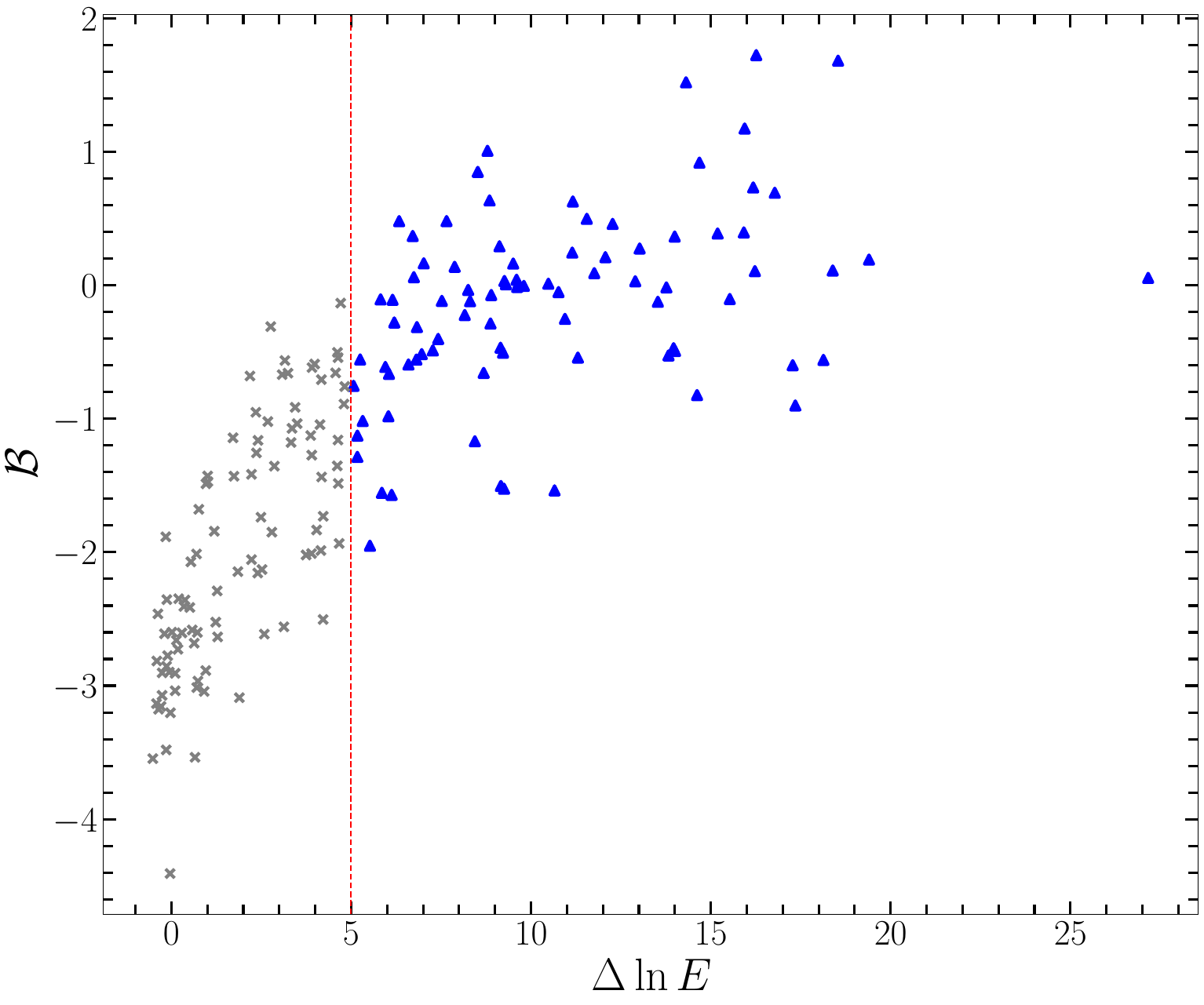}
    \caption{Relation between the mass error coefficient $\mathcal{B}$ and the logarithmic evidence increase $\Delta \ln E$. Blue triangles represent significant detections with $\Delta \ln E > 5$. Grey crosses represent tentative detections with only $\Delta \ln L > 10$. Under the current detection threshold (red dashed line), no apparent correlation is observed between $\mathcal{B}$ and $\Delta \ln E$. This suggests that a logarithmic evidence increase of $\Delta \ln E > 5$ is sufficient to obtain unbiased individual subhalo measurements using Euclid strong lensing data.}
    \label{fig:bias_evidence}
\end{figure}

Using a detection threshold of $\Delta\ln E > 5$, we obtain 86 detections. Figure~\ref{fig:bias_position} shows the distribution of the error coefficient $\mathcal{B}$ for these detected subhalos. Their spatial distribution agrees with the sensitivity map of \citet{Euclid_sensitivity}, with most detections occurring near the brightest regions of the lensed arc. The subhalo mass can be accurately recovered regardless of its location, provided it meets our detection criteria of $\Delta\ln E>5$. 
To further investigate the robustness of our fitting results, we plot the relationship between the error coefficient and the Bayesian evidence increase in Figure~\ref{fig:bias_evidence}. For the 86 detections, we find no correlation between $\mathcal{B}$ and $\Delta\ln{E}$. In most cases, $M_{200}$ is recovered within $68\%$ credible interval (i.e. $|\mathcal{B}|<1$). 
We therefore conclude that a detection threshold of $\Delta\ln E>5$ is sufficient to achieve unbiased subhalo mass measurements for Euclid-like data. For comparison, Figure~\ref{fig:bias_evidence} also shows results for samples that do not meet our evidence-based detection threshold but have an increase in the maximum log-likelihood of 10 or larger, which could be considered as \textit{tentative detections} \citep{He2023}. However, they present a notable underestimation of their masses and a strong positive correlation between $\mathcal{B}$ and $\Delta\ln{E}$. This result underscores the importance of using Bayesian evidence, rather than likelihood-based metrics, to establish a robust subhalo detection threshold.


\begin{figure*}
    \centering
    \begin{minipage}[t]{0.48\linewidth}
        \includegraphics[width=\linewidth]{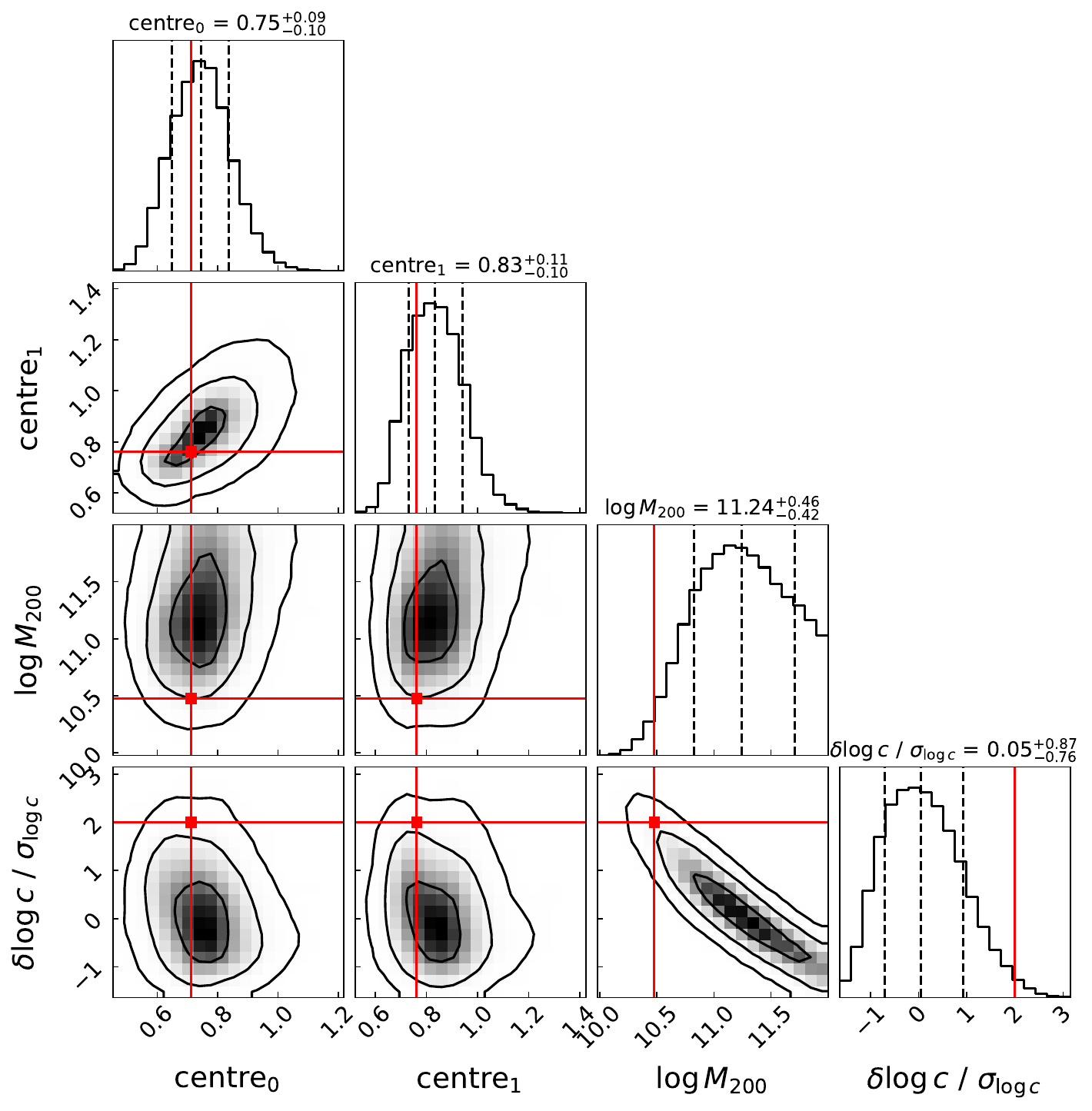}
    \end{minipage}
    \begin{minipage}[t]{0.48\linewidth}
        \includegraphics[width=\linewidth]{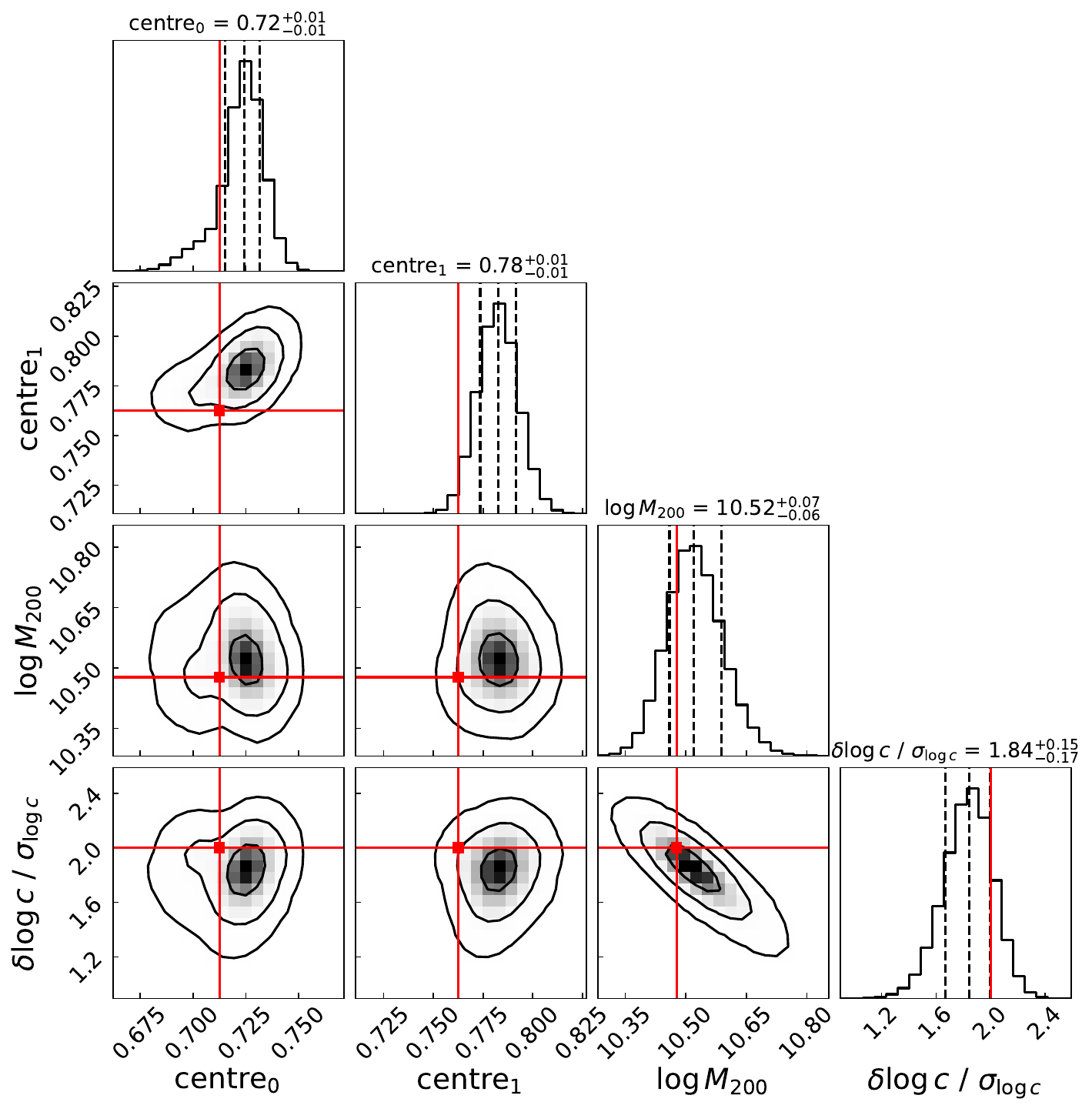}
    \end{minipage}
    \caption{Posterior distributions of model parameters for a simulated subhalo derived from lensing data with different angular resolutions: Euclid (left) and HST (right). The true subhalo concentration is $2\sigma$ above the median \citet{mcr_Ludlow} relation. True parameter values are indicated by red lines. The 2D contours denote the $68\%$, $95\%$, $99\%$ credible regions. In the 1D marginalised histograms, the $68\%$ credible interval is shown by grey dashed lines. We treat the subhalo concentration as a free parameter by sampling its normalised offset from the \citet{mcr_Ludlow} relation, $s_{\rm c}\equiv \delta\log c / \sigma_{\log c}$, where $\sigma_{\log c}=0.15$.}
    \label{fig:corner}
\end{figure*}

In the real Universe, intrinsic scatter in the mass-concentration relation produces a diversity of subhalo profiles, from compact to diffuse. Owing to the degeneracy between subhalo mass and concentration in strong lens modelling \citep{concentration_Amorisco2021}, it remains uncertain whether both properties can be measured simultaneously and accurately with Euclid-quality lensing data. To investigate this, we simulate a Euclid-like mock lens containing an NFW subhalo with a concentration $2\sigma$ above the median L16 relation. Fitting this image yields the posterior distribution shown in the left panel of Figure~\ref{fig:corner}. In the $(M_{200}, s_{\rm c})$ plane, the contour exhibits the characteristic banana shape, demonstrating the degeneracy between these two parameters. Along this contour, increasing concentration while decreasing mass produces similar lensing perturbations. Although both parameters are recovered within the $99\%$ credible interval, the corresponding uncertainty in $M_{200}$ spans nearly an order of magnitude. Without prior knowledge of the true subhalo profile, the injected compact subhalo would be misidentified as a more massive ($M_{200}\sim10^{11}\,{\rm M}_\odot$) subhalo lying on the median L16 relation (with $s_{\rm c} = \delta\log c / \sigma_{\log c}\sim0$). As a complementary test, we maintain the same SNR in the lensing image but increase the spatial resolution from Euclid (PSF FWHM = $0.18\arcsec$, pixel scale = $0.1\arcsec$) to HST (PSF FWHM = $0.09\arcsec$, pixel scale = $0.05\arcsec$). We find both quantities to be accurately measured within the 68\% credible interval (see the right-hand panel of Figure~\ref{fig:corner}). We conclude that the spatial resolution of Euclid is insufficient to fully break the subhalo mass-concentration degeneracy in strong-lensing analyses, preventing unbiased measurements of both quantities. Achieving this requires the higher resolution provided by facilities such as HST.

\subsection{Assessing the impact of satellite galaxies hosted by subhalos}
\label{sec:galaxy}
A subhalo with mass $10^{10}\,{\rm M}_\odot$ may host a dwarf satellite galaxy whose stellar mass and light distribution can affect the accuracy of halo-mass measurements. We assess the impact of such putative satellites on lens modelling in two stages. In Section~\ref{sec:sat_gal_bias}, we first test whether satellite galaxies hosted by subhalos bias subhalo-mass estimates when not explicitly included in the lens model. If a bias is present, we then assess whether it can be mitigated by explicitly including the satellite-galaxy components in the lens model (Section~\ref{sec:model_sat_gal}).

\subsubsection{Bias from the satellite galaxy}
\label{sec:sat_gal_bias}

We assess whether the presence of a satellite galaxy hosted by the subhalo biases the subhalo mass measurement in our lensing analyses. To this end, we simulate four datasets, each comprising 100 Euclid-like lensing images. Within each dataset, all physical parameters used to generate the mock images are identical; only the noise realisation varies. Specifically, the four datasets are: (1) a control dataset with only an NFW subhalo perturber; (2) a dataset including the NFW subhalo plus the stellar-mass component of the putative satellite galaxy it hosts; (3) a dataset including the NFW subhalo plus the stellar light component of the putative satellite galaxy; and (4) a dataset including the NFW subhalo plus both the stellar-mass and light components of the putative satellite galaxy. We adopt a stellar mass-to-light ratio of $2\,{\rm M}_\odot/{\rm L}_\odot$ for the putative satellite and assume that the satellite follows the SHMR of \citet{Girelli2020}. The parameter values used to generate the mock images are summarised in Table~\ref{tab:param}. Our aim is to test whether the subhalo mass, specifically $M_{200}$, can be recovered without bias from strong-lensing data when the internal structure of the subhalo (for example, whether it hosts a satellite galaxy) is not known \emph{a priori}, as is typically the case for real lenses. Accordingly, we always model the subhalo using a single NFW profile,\footnote{For subhalos that are intrinsically NFW+S\'ersic in the simulations, the target quantity we seek to recover remains the subhalo's $M_{200}$, defined with respect to the NFW halo.} even when the underlying simulated subhalo includes additional stellar components.

\begin{figure}
    \centering
    \includegraphics[width=\columnwidth]{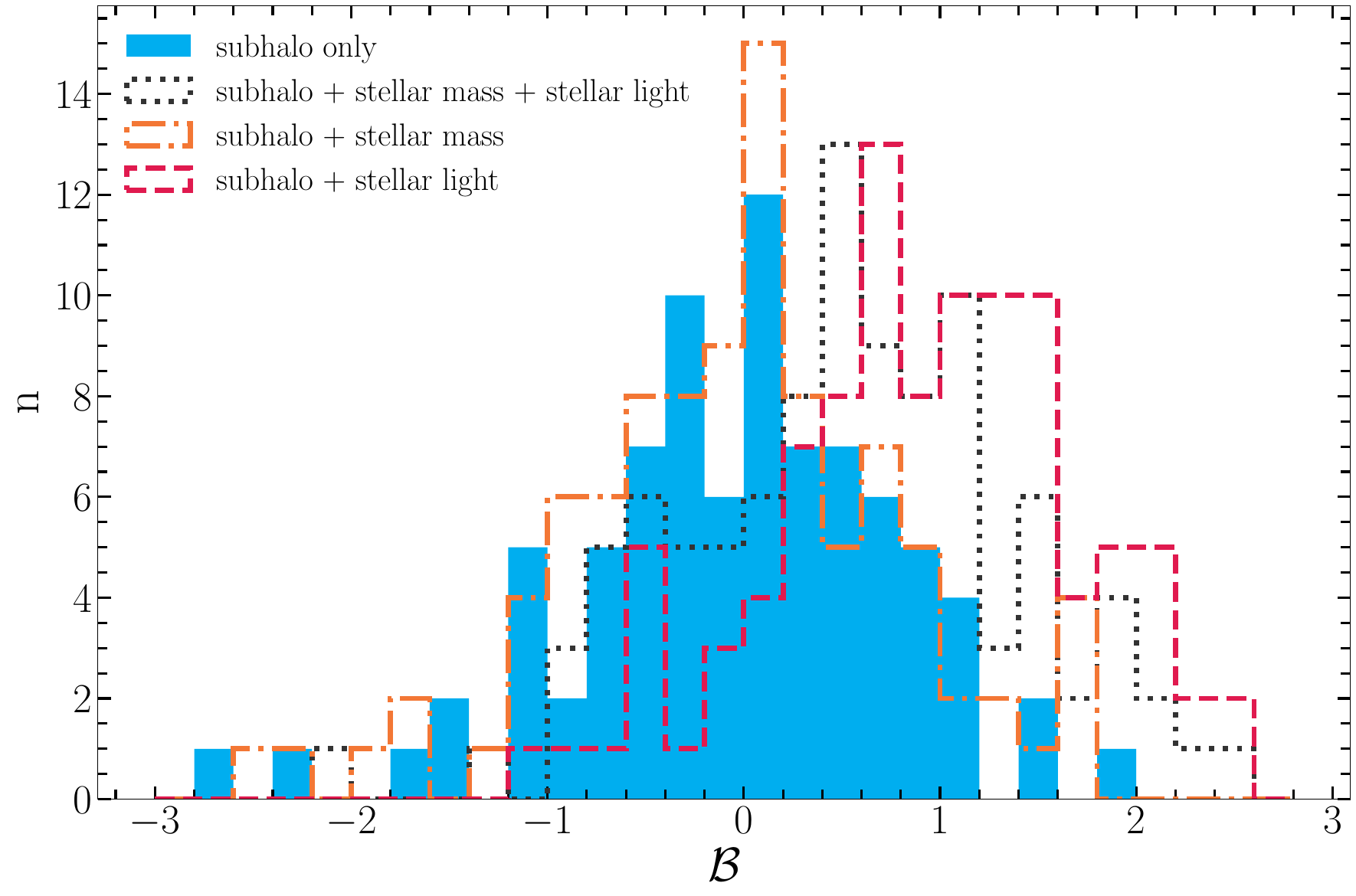}
    \caption{Histograms of the mass-error coefficient, $\mathcal{B}$, for four simulated datasets, each comprising 100 lensing images. Within each dataset, all mock images share identical physical parameters and differ only in their noise realisations. The datasets are: (1) lenses with an NFW subhalo only (blue histogram); (2) lenses with an NFW subhalo plus the satellite galaxy’s stellar-mass component (orange dot–dashed); (3) lenses with an NFW subhalo plus the satellite galaxy’s stellar-light component (red dashed); and (4) lenses with an NFW subhalo plus both the satellite galaxy’s stellar-mass and stellar-light components (black dotted).}
    \label{fig:hist}
\end{figure}

Figure~\ref{fig:hist} presents the distribution of the mass error coefficient $\mathcal{B}$ for different mock datasets. Control samples without a satellite galaxy show no systematic bias (blue histogram), confirming the robustness of our pipeline for standard NFW subhalos. By contrast, samples that include both the stellar-mass and light contributions of satellite galaxies hosted by the subhalo yield a pronounced bias, with a mean error coefficient $\bar{\mathcal{B}}=0.55$ (black dotted histogram), indicating that unmodelled satellite emission affects the inferred subhalo mass. To isolate the cause, we find that adding only the satellite’s stellar-mass component remains unbiased (orange dot–dashed histogram), implying that the unmodelled stellar mass alone does not induce a significant bias. Conversely, including only the satellite’s light produces a substantial overestimation of the subhalo mass, with $\bar{\mathcal{B}}=0.92$ (red dashed histogram), demonstrating that the systematic shift is driven by the stellar light. This behaviour likely arises because satellite light is partly misattributed to the lensed background source, artificially increasing the lensing perturbation attributed to the subhalo. Although individual subhalo masses remain consistent with the true values within their $99\%$ credible intervals, the population-level bias evident in Figure~\ref{fig:hist} can be important when stacking measurements to constrain the SHMR.

\subsubsection{Modelling the satellite galaxy}
\label{sec:model_sat_gal}
\begin{figure}
    \centering
    \includegraphics[width=\columnwidth]{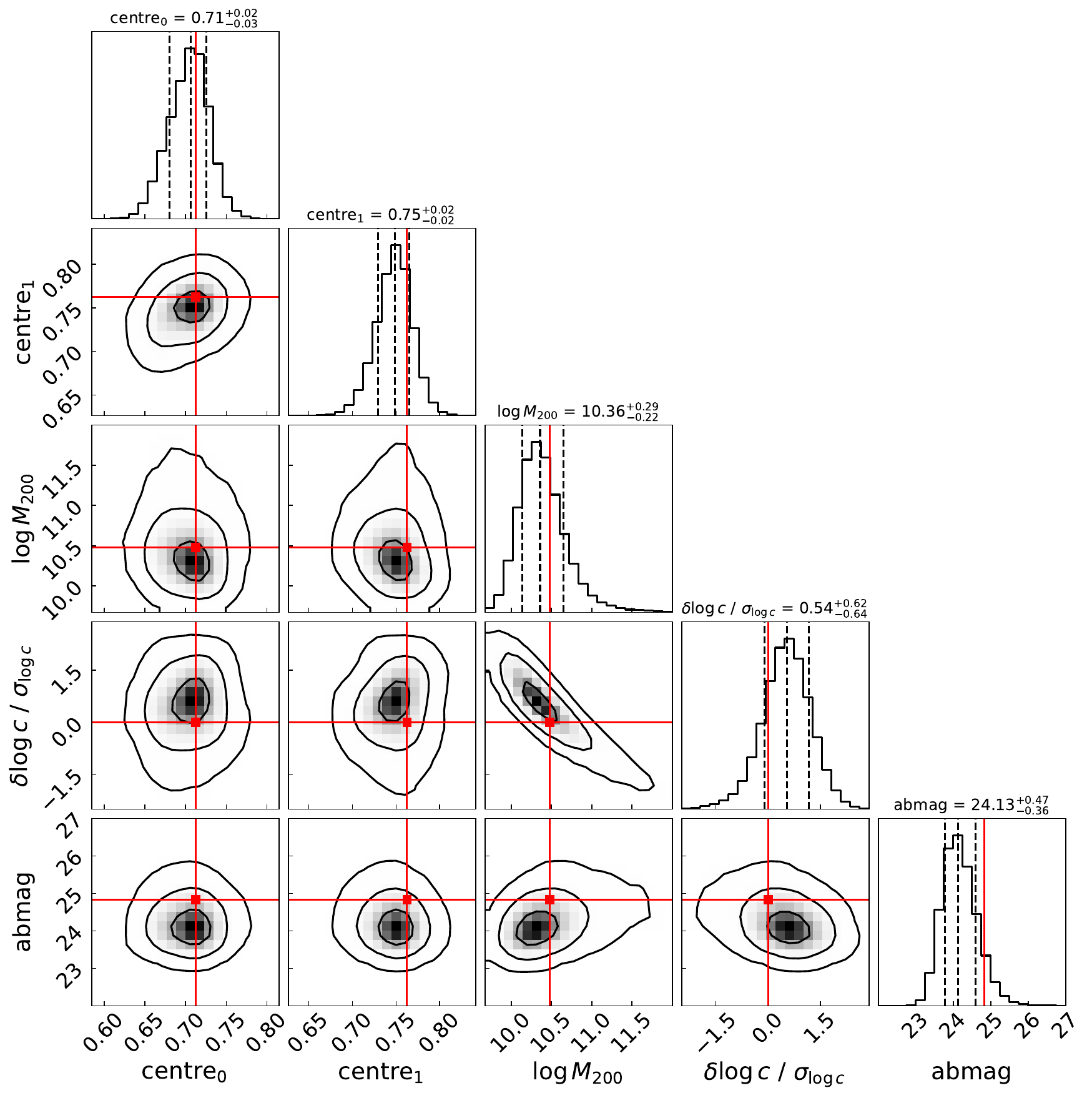}
    \caption{Posterior distribution for the model parameters of the simulated subhalo when simultaneously fitting the subhalo mass and the satellite light. Both $M_{200}$ and satellite magnitude are simultaneously recovered within $95\%$ credible interval.}
    \label{fig:galaxy_reconstruction}
\end{figure}

As established in Section~\ref{sec:sat_gal_bias}, ignoring a satellite galaxy's light biases the mass measurement of its host subhalo. We therefore investigate whether explicitly including the satellite's light in the lensing model enables accurate measurement of both the galaxy's magnitude and the subhalo's mass properties, following the work of \citet{he2025darkdensealternativeexplanation}. We find that lensing data at the resolution of \textit{Euclid} are insufficient to recover the satellite's apparent magnitude and the subhalo's mass and concentration simultaneously. The apparent magnitude can be underestimated by $\sim0.2$~mag, whilst the subhalo mass is overestimated by $\sim1$~dex and its concentration is underestimated by $\sim$ 50\%. However, if the image resolution is increased to that of \textit{HST}, we find that the satellite's apparent magnitude, the subhalo's mass, and its concentration can all be recovered accurately and within the  68\% credible interval. This is demonstrated by the posterior distribution for an example lens system shown in Figure~\ref{fig:galaxy_reconstruction}. We therefore conclude that to measure accurately both the satellite galaxy's magnitude and its host subhalo's properties, the image resolution of \textit{Euclid} is insufficient, and high-resolution follow-up observations from facilities such as \textit{HST} are required.

\subsection{Constraining stellar-to-halo mass relation}
\label{sec:constrain}
Strong lensing enables direct measurements of subhalo masses\footnote{In this work, we use $M_{200}$ as a proxy for the halo mass $M_{\rm h}$. This choice may be an oversimplification, because subhalos can undergo substantial tidal stripping and lose mass after accretion into the main halo. Future studies could adopt a truncated NFW profile to provide a more realistic representation of subhalos.} at cosmological distances. Combined with an independent estimate of the hosted satellite galaxy’s stellar mass, derived from stellar population synthesis (SPS) modelling of photometric spectral energy distributions (SEDs), these measurements permit constraints on the SHMR.

\begin{figure}
    \centering
    \includegraphics[width=\columnwidth]{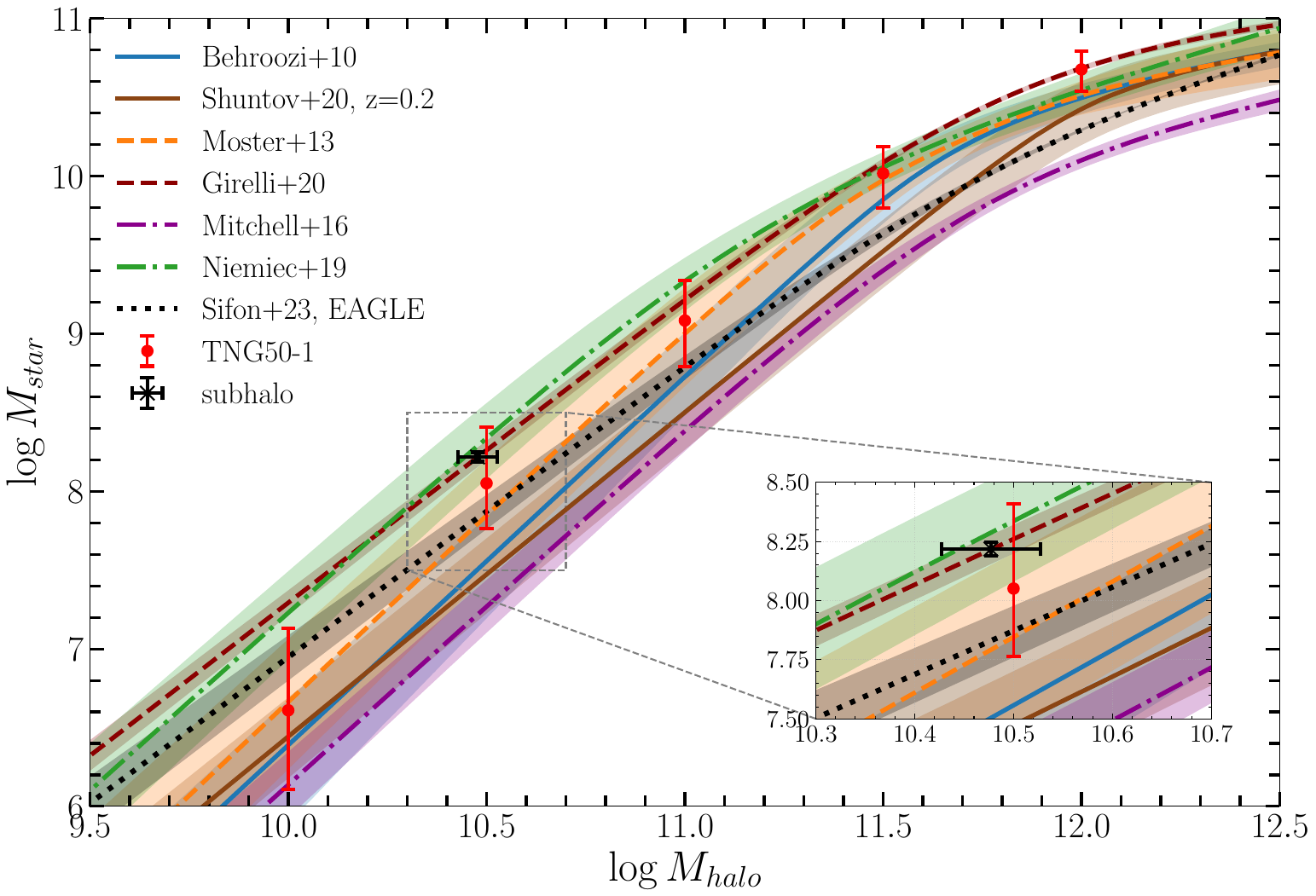}
    \caption{
The mean SHMR at a halo mass of $\sim 10^{10}\,{\rm M}_\odot$ can be constrained by statistically stacking the stellar and halo mass measurements. Stellar masses are derived from SPS modelling, while halo masses are determined through strong lensing analysis. The input subhalos are drawn from the fiducial SHMR of \citet{Girelli2020} (orange dashed line). Assuming we have 100 subhalo detections of $\sim10^{10}{\rm M}_\odot$ as presented in this work, the result constraint is shown as the brown marker; its horizontal and vertical error bars indicate the uncertainties in the stacked halo and stellar masses, respectively. For comparison, alternative SHMR models are shown as other lines, with shaded regions indicating the $68\%$ uncertainty intervals. The red crosses with error bars indicate the median stellar mass and the $16^{\rm th}\sim 84^{\rm th}$ percentile scatter for different halo mass bins in TNG50 simulation.
    }
    \label{fig:shmr}
\end{figure}

Our mock lensing dataset comprises dwarf galaxies with stellar masses of $\sim 10^{8}\,{\rm M}_\odot$ and halo masses of $\sim 10^{10}\,{\rm M}_\odot$. Based on the results of our mock tests,\footnote{As shown in Section~\ref{sec:galaxy}, the image resolution of \textit{Euclid} is insufficient to recover the subhalo light and mass simultaneously. Therefore, the discussion in this section applies to mock lenses simulated at \textit{HST} resolution.} we assume that the stellar mass of each dwarf galaxy is measured without bias, with a typical uncertainty of $\sim 0.3$~dex \citep{SPS_err_1, SPS_err_2, SPS_err_3}. Furthermore, as demonstrated in Section~\ref{sec:galaxy}, by incorporating both the mass and light components into the lensing modelling, it is possible to obtain an unbiased measurement of the subhalo's total mass, with a typical uncertainty of $\sim 0.5$ dex. To constrain the SHMR, we stack the individual measurements using an inverse-variance weighted mean:
\begin{equation}
M = \frac{\sum\limits_{i} M_i \, \sigma_i^{-2}}{\sum\limits_{i} \sigma_i^{-2}},\quad
\sigma^2 = \frac{1}{\sum\limits_{i} \sigma_i^{-2}},
\end{equation}
where $M_i$ and $\sigma_i$ denote the mass (either stellar or halo) and its corresponding uncertainty for dwarf galaxy $i$. We predict that a statistical sample of 100 subhalos discovered with \textit{Euclid} will place a robust constraint on the SHMR, with average uncertainties of $\sim 0.03$ and $\sim 0.05$ dex for the stellar and halo mass, respectively. This level of precision is sufficient to distinguish statistically between competing SHMR predictions derived from different methods and assumptions, as shown in Figure~\ref{fig:shmr}

\section{Discussion}
\label{sec:discussion}
\subsection{Simplifications and caveats in the simulations}
\label{sec:caveats}
We adopted several simplifications in lensing simulations that could affect the accuracy of subhalo mass measurements. Although we expect these not to alter our main conclusions, we outline them below for completeness.

First, we do not systematically investigate how the accuracy of subhalo mass measurements depends on the lensing geometry and image morphology; we show only that, in regions where the subhalo is most sensitively detected (near the Einstein ring), the subhalo mass can be measured accurately when the detection significance is high. Second, our mock neglects lens light, assumes a simple S\'ersic source, and excludes additional angular mass complexity (e.g. boxiness and disciness) that can be partly degenerate with the subhalo signal; these simplifications facilitate the measurement of subhalo properties. Nevertheless, we expect the generality of our conclusions to remain robust, because advanced lens-modelling techniques---such as multi-Gaussian expansion (MGE) lens-light modelling \citep{He2024}, adaptive brightness-based pixelised source models \citep[e.g.][]{he2025darkdensealternativeexplanation}, and multipole mass components to account for boxiness and disciness \citep[e.g.][]{Nightingale2024, Lange2024}---can accommodate the complexity present in real observational data and yield accurate subhalo measurements.

\subsection{Can we measure stellar masses accurately?}
\label{sec:stellar}
For a subhalo with a mass of $\sim 10^{10}$~${\rm M}_\odot$, various theoretical SHMRs predict a corresponding stellar mass in the range $10^{7.2}$--$10^{8.3}$~${\rm M}_\odot$. Based on the typical stellar mass-to-light ratios of dwarf galaxies \citep{Du2020}, such a subhalo is expected to host a dwarf galaxy with a luminosity of $10^{7.6}$--$10^{8.7}$~L$_{\odot}$. Assuming the main lens galaxy is at a redshift of 0.2, a typical value for the well-known SLACS strong lens sample \citep{SLACSV}, the apparent magnitude of such a dwarf galaxy would be approximately $23.2$--$25.9$. For leading facilities such as the James Webb Space Telescope (JWST) or next-generation instruments like the Extremely Large Telescope (ELT), obtaining accurate multi-band photometry for such dwarf galaxies is feasible. This enables reliable stellar mass measurements to be derived through spectral energy distribution (SED) fitting. A further complication is that the light from the dwarf satellite galaxy cannot typically be measured directly, as it is usually embedded within the emission of the lensed source. This means that the photometry of the dwarf satellite must be deblended using the lens model. Fortunately, results from \cite{he2025darkdensealternativeexplanation} and tests conducted in this study demonstrate that such a deblending is achievable. In summary, we expect that the stellar mass of dwarf satellite galaxies within $\sim$10$^{10}$ ${\rm M}_\odot$ subhalos can be robustly measured using SED fitting techniques with data from premier telescope facilities.

\subsection{Hierarchical Bayesian inference of the SHMR}
\label{sec:HBayesian}
In Section~\ref{sec:constrain}, we stacked a sample of strong-lensing systems with subhalo detections around $\sim 10^{10}\,{\rm M}_\odot$ to illustrate the statistical constraining power on the SHMR at this halo-mass scale. However, that procedure is deliberately simplified and serves only as a conceptual demonstration. A more rigorous framework for statistically constraining the SHMR is hierarchical Bayesian inference, which enables robust estimation of the underlying SHMR parameters by self-consistently modelling observational uncertainties and intrinsic population scatter. We outline this framework below.

We parameterise the local SHMR as a linear relation in logarithmic space, an effective approximation over a narrow range of halo masses. The expected stellar mass, $m_{*} \equiv \log_{10}(M_{*} / {\rm M}_\odot)$, for a given true halo mass, $m_{\rm h,true} \equiv \log_{10}(M_{\rm h,true} / {\rm M}_\odot)$, is
\begin{equation}
\mathbb{E}[m_{*} \mid m_{\rm h,true}] = \alpha + \beta \left(m_{\rm h,true} - m_{\rm h,piv}\right),
\label{eq:shmr_mean}
\end{equation}
where $\alpha$ is the mean stellar mass at the pivot halo mass and $\beta$ is the local slope. For numerical stability and to reduce covariance between $\alpha$ and $\beta$, we set the pivot mass to the mean of the observed logarithmic halo masses, i.e. $m_{\rm h,piv} = \mathbb{E}[m_{{\rm h,obs},i}]$.

Equation~\eqref{eq:shmr_mean} describes the mean trend, but individual subhalos are expected to scatter around this relation owing to diverse formation histories. We model this intrinsic scatter, $\sigma_{\rm int}$, by assuming that true logarithmic stellar masses are drawn from a Gaussian distribution about the mean relation:
\begin{equation}
m_{{\rm *,true}} \sim \mathcal{N}\!\left(\mathbb{E}[m_{*} \mid m_{\rm h,true}], \sigma_{\rm int}^2\right),
\label{eq:intrinsic_scatter}
\end{equation}
where $\mathcal{N}(\mu, \sigma^2)$ denotes a Gaussian distribution with mean $\mu$ and variance $\sigma^2$.

The true masses, $m_{\rm h, true}$ and $m_{\rm *, true}$, are not directly observable. Instead, for each strong-lensing system $i$ in our sample of $N$ systems, we have noisy measurements, $m_{\rm h, obs, i}$ and $m_{\rm *, obs, i}$, with associated measurement uncertainties, $\sigma_{\rm h, obs, i}$ and $\sigma_{\rm *, obs, i}$, respectively. We assume these observational errors are Gaussian and independent, linking the observed values to the true (latent) values as follows:
\begin{align}
m_{\rm h, obs, i} &\sim \mathcal{N}(m_{\rm h, true, i}, \sigma_{\rm h, obs, i}^2), \label{eq:obs_h} \\
m_{\rm *, obs, i} &\sim \mathcal{N}(m_{\rm *, true, i}, \sigma_{\rm *, obs, i}^2). \label{eq:obs_star}
\end{align}

The complete hierarchical model is defined by Equations~\eqref{eq:shmr_mean}--\eqref{eq:obs_star}. Our goal is to infer the posterior distribution of the population-level hyperparameters, $\Theta = \{\alpha, \beta, \sigma_{\rm int}\}$, given the observed data set $\mathcal{D} = \{m_{{\rm h,obs},i}, m_{{\rm *,obs},i}, \sigma_{{\rm h,obs},i}, \sigma_{{\rm *,obs},i}\}_{i=1}^N$. Within the Bayesian framework, the unknown true masses for each system, $\{m_{{\rm h,true},i}, m_{{\rm *,true},i}\}$, are treated as latent variables and marginalised over. To complete the model, we adopt uninformative (or weakly informative) priors for the hyperparameters. By simultaneously modelling the underlying physical relation, its intrinsic scatter, and measurement errors, this framework propagates all sources of uncertainty into the final constraints on the SHMR parameters.

We infer the joint posterior of all parameters using Markov chain Monte Carlo (MCMC). Our implementation employs the No-U-Turn Sampler (NUTS; \citealt{NUTS}) within the \texttt{NumPyro} probabilistic programming library \citep{Numpyro}. To ensure a robust exploration of the parameter space, we run multiple independent chains and confirm convergence using the Gelman–Rubin diagnostic ($\hat{R}$). Applying this framework to our mock data, we constrain the mean logarithmic stellar mass, $\alpha$, to a precision of $\sim 0.06$\,dex at a pivot halo mass of $m_{\rm h,piv}=\log_{10}(3 \times 10^{10})$.

\subsection{Revisiting imaging resolution}
\label{sec:resolution}

As established in Sections~\ref{sec:position} and \ref{sec:sat_gal_bias}, the angular resolution of \textit{Euclid} is insufficient for robustly measuring subhalo masses, owing to the mass--concentration degeneracy and the difficulty in deblending the satellite's light from the mass-induced perturbation. Beyond the accuracy of mass measurements, image resolution is also critical for establishing the statistical significance of subhalo detections. 

Real gravitational lenses can exhibit complex mass distributions or other systematics that are not fully captured by idealised lens models. These unmodelled features can mimic the lensing signal of a subhalo, leading to potential false-positive detections. Consequently, a stringent detection threshold is required in practice, such as a Bayesian evidence increase of $\Delta \ln E > 10$ \citep{Nightingale2024} or even $\Delta \ln E > 50$ \citep{Despali2024}. In our simulations, Euclid-like data rarely reach these levels: of the 86 subhalos flagged as detected in Section~\ref{sec:position}, only 36 satisfy $\Delta \ln E > 10$, and only one exceeds 20. Thus, a substantial fraction of Euclid candidates would remain ambiguous under evidence criteria appropriate for real lenses.

\begin{figure}
    \centering
    \includegraphics[width=\columnwidth]{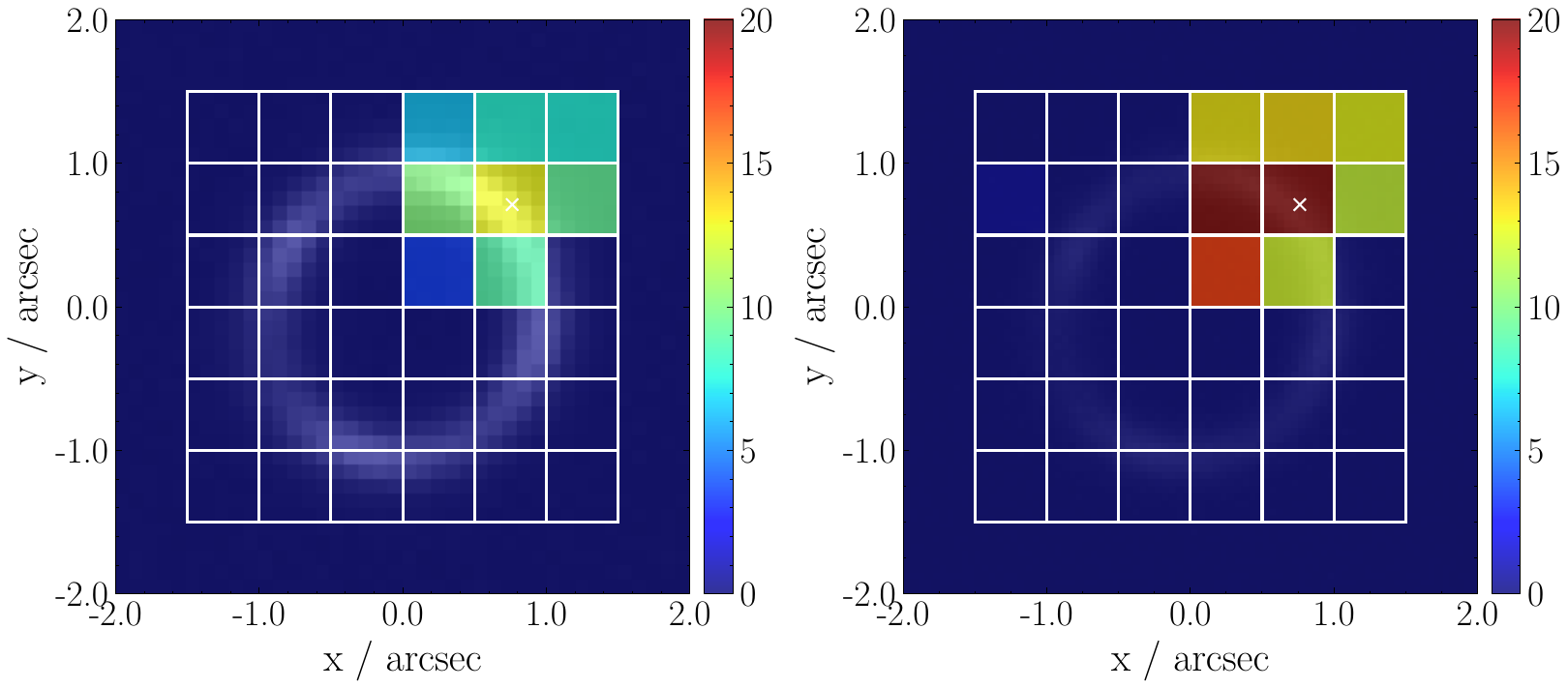}
    \caption{Analogous to Figure~\ref{fig:example_detection}, we compare evidence increase maps from the grid search for two lensing images that differ only in spatial resolution. The colour scale indicates $\Delta \ln E$, and white crosses mark the input subhalo position. \textit{Left:} Euclid-like resolution (PSF FWHM $= 0.18\arcsec$, pixel scale $= 0.1\arcsec$). \textit{Right:} HST-like resolution (PSF FWHM $= 0.09\arcsec$, pixel scale $= 0.05\arcsec$).}
    \label{fig:resolution}
\end{figure}

High-resolution imaging mitigates these limitations by resolving finer structure in the arcs, thereby increasing sensitivity to the small-scale distortions induced by subhalos and boosting the Bayesian evidence for perturbative models. To illustrate this, we simulate two lensing images that differ only in angular resolution—one Euclid-like and one HST-like—and analyse both with the same modelling pipeline. For the Euclid-like case, the subhalo model improves the evidence by $\Delta \ln E = 12.82$; for the HST-like case, the improvement increases to $\Delta \ln E = 20.07$. The corresponding enhancement is visible in the grid search shown in Figure~\ref{fig:resolution}. These results underscore the necessity of deep, high-resolution follow-up with facilities such as HST and JWST. Such observations will both increase detection significance and help to break modelling degeneracies, enabling the confident confirmation or rejection of the many low-significance subhalo candidates expected from Euclid and, ultimately, a more robust determination of the SHMR at dwarf-galaxy scales.

\section{Conclusion}
\label{sec:conclusion}
In this work, we have conducted an end-to-end analysis to assess the capability of the new generation of space-based imaging surveys, specifically \textit{Euclid}, to constrain the stellar-to-halo mass relation (SHMR) at the dwarf galaxy scale ($M_{\rm h} \sim 10^{10}\,{\rm M}_\odot$). Using a suite of simulated galaxy-galaxy strong lensing observations tailored to the specifications of the \textit{Euclid} VIS instrument, we systematically quantified the accuracy and potential biases in recovering dark-matter subhalo masses, and examined how subhalo properties (position and concentration) and baryonic components of hosted satellites impact those inferences.

Our main conclusions are as follows.
\begin{enumerate}
    \item For an idealised NFW subhalo with $M_{\rm h}\sim 10^{10}\,{\rm M}_\odot$ and concentration set by the standard mass--concentration relation, \textit{Euclid}-quality imaging enables an unbiased mass recovery, provided the detection is significant (evidence increase $\Delta\ln E \geq 5$, i.e. $>3.6\sigma$).

    \item In realistic populations, subhalos at fixed mass span a range of concentrations, requiring simultaneous inference of mass and concentration. At \textit{Euclid}'s angular resolution, the intrinsic mass--concentration degeneracy cannot be broken: the 99\% credible interval for concentration can span nearly an order of magnitude, such that a compact, low-mass subhalo may be misinterpreted as a more massive, less concentrated one lying on the standard mass-concentration relation.

    \item A $M_{\rm h}\sim 10^{10}\,{\rm M}_\odot$ subhalo is expected to host a satellite galaxy with $M_{\rm *}\sim 10^{8}\,{\rm M}_\odot$ and $L_{\rm *}\sim 10^{8}\,{\rm L}_\odot$. If the satellite light is not explicitly modelled, it induces a systematic overestimate of the subhalo mass, with an average bias of $0.92\sigma$. By contrast, neglecting the satellite's stellar mass component alone does not introduce a comparable bias.

    \item High-resolution follow-up imaging is essential. Data of HST quality can mitigate the mass-concentration degeneracy and allow for the satellite's light to be modelled simultaneously, enabling an accurate recovery of the halo mass, concentration, and satellite magnitude. Higher resolution also substantially increases the Bayesian evidence for perturbative models (e.g. boosting $\Delta\ln E$ from $\simeq 13$ to $\simeq 20$ in our tests), which is critical for confirming the numerous low-significance subhalo candidates expected from \textit{Euclid}.

    \item By combining lensing-derived halo masses with stellar masses from the spectral energy distribution (SED) fitting of multi-band satellite photometry (also derived from the lens model), a statistical sample of $\sim 100$ detections can deliver powerful constraints on the SHMR at $M_{\rm h}\sim 10^{10}\,{\rm M}_\odot$, achieving characteristic uncertainties of $\simeq 0.05$\,dex (halo mass) and $\simeq 0.03$\,dex (stellar mass). This precision suffices to discriminate among competing SHMR models at dwarf-galaxy scales.

\end{enumerate}

In summary, wide-field space surveys such as \textit{Euclid} and the CSST will discover an unprecedented number of strong lenses hosting dwarf-scale subhalos. Realising the full scientific return requires a synergistic strategy that couples survey discovery with targeted, high-resolution follow-up (e.g. HST/JWST), enabling robust deblending of satellite light, mitigation of mass-concentration degeneracies, and high-significance subhalo confirmations. This approach will unlock strong gravitational lensing as a precise probe of the galaxy--halo connection at the low-mass frontier.

\section*{Acknowledgements}

We acknowledge the support by National Key R\&D Program of China No. 2022YFF0503403, the support of National Nature Science Foundation of China (No. 11988101), the support from the Ministry of Science and Technology of China (No. 2020SKA0110100), the science research grants from the China Manned Space Program (No. CMS-CSST-2025-A03), CAS Project for Young Scientists in Basic Research (No. YSBR-062), and the support from K.C.Wong Education Foundation. XYC acknowledges the support of the National Natural Science Foundation of China (No. 12303006). This work was performed using the Cambridge Service for Data Driven Discovery (CSD3), part of which is operated by the University of Cambridge Research Computing on behalf of the STFC DiRAC HPC Facility (www.dirac.ac.uk). The DiRAC component of CSD3 was funded by BEIS capital funding via STFC capital grants ST/P002307/1 and ST/R002452/1 and STFC operations grant ST/R00689X/1. DiRAC is part of the National e-Infrastructure. This work also used the DiRAC@Durham facility managed by the Institute for Computational Cosmology on behalf of the STFC DiRAC HPC Facility (www.dirac.ac.uk). The equipment was funded by BEIS capital funding via STFC capital grants ST/P002293/1 and ST/R002371/1, Durham University and STFC operations grant ST/R000832/1. DiRAC is part of the National e-Infrastructure.

%%%%%%%%%%%%%%%%%%%%%%%%%%%%%%%%%%%%%%%%%%%%%%%%%%
\section*{Data Availability}
 
The software used for lens modelling, {\tt PyAutoLens}, is open source and available at \url{https://github.com/Jammy2211/PyAutoLens}. Model parameters are available in Table~\ref{tab:param}. Simulated images and fitting results are available upon request; please contact the authors.



\bibliographystyle{mnras}
\bibliography{reference}
% \bibliography{reference, james} % if your bibtex file is called example.bib

%\begin{thebibliography}{99}
%\end{thebibliography}

%%%%%%%%%%%%%%%%%%%%%%%%%%%%%%%%%%%%%%%%%%%%%%%%%%

%%%%%%%%%%%%%%%%% APPENDICES %%%%%%%%%%%%%%%%%%%%%

%\appendix

%%%%%%%%%%%%%%%%%%%%%%%%%%%%%%%%%%%%%%%%%%%%%%%%%%

% Don't change these lines
\bsp	% typesetting comment
\label{lastpage}
\end{document}